\documentclass[final,5p,times,twocolumn]{elsarticle}

\usepackage{graphics}
\usepackage{tabto}
\usepackage{bm}
\usepackage{amsmath}
\usepackage{wrapfig,lipsum,booktabs}
\usepackage{amssymb}
\usepackage{url}
\usepackage{color}
\usepackage{titlesec}
\usepackage{mathtools, cuted}
\usepackage{widetext}

\newcounter{bla}

\journal{Computer Physics Communications}
\begin{document}
\begin{frontmatter}

\title{A generalized massively parallel ultra-high order FFT-based Maxwell solver}

\author[1]{Haithem Kallala\thanks{haithem.kallala@cea.fr}}
\author[2]{Jean-Luc Vay\thanks{jlvay@lbl.gov}}
\author[3]{Henri Vincenti\thanks{henri.vincenti@cea.fr}\corref{author}}

\cortext[author] {Corresponding author.\\\textit{E-mail address:} henri.vincenti@cea.fr}
\address[1]{Maison de la Simulation, CEA, CNRS, Universit\'e Paris-Saclay, CEA Saclay, 91 191 Gif-sur-Yvette, France}
\address[2]{Lawrence Berkeley National Laboratory, Berkeley, CA, USA}
\address[3]{LIDYL, CEA, CNRS, Universit\'e Paris-Saclay, CEA Saclay, 91 191 Gif-sur-Yvette, France}


\begin{abstract}
Dispersion-free ultra-high order FFT-based Maxwell solvers have recently proven to be paramount to a large range of applications, including the high-fidelity modeling of high-intensity laser-matter interactions with Particle-In-Cell (PIC) codes. To enable a massively parallel scaling of these solvers, a novel parallelization technique was recently proposed, which consists in splitting the simulation domain into several processor sub-domains, with guard regions appended at each sub-domain boundaries. Maxwell's equations  are advanced independently on each sub-domain using local shared-memory FFTs (instead of a single distributed global FFT). This implies small truncation errors at sub-domain boundaries, the amplitude of which depends on guard regions sizes and order of the Maxwell solver. For moderate guard region sizes, this 'local' technique proved to be highly scalable on up to a million cores and notably enabled the 3D modelling of so-called plasma mirrors, for which $8$ guard cells only were enough to prevent truncation error growth. Yet, for other applications, the required number of guard cells might be much higher, which would severely limit the parallel efficiency of this technique due to the large volume of guard cells to be exchanged between sub-domains. In this context, we propose a novel parallelization technique that ensures very good scaling of FFT-based solvers with an arbitrarily high number of guard cells. Our 'hybrid' technique consists in performing distributed FFTs on local groups of processors with guard regions now appended to boundaries of each group of processors. It uses a dual domain decomposition method for the Maxwell solver and other parts of the PIC cycle to keep the simulation load-balanced.  This 'hybrid' technique was implemented in the open source exascale library PICSAR. Benchmarks show that for a large number of guard cells ($>16$), the 'hybrid' technique offers a $\times 3$ speed-up and $\times 8$ memory savings compared to the 'local' one.
%
\end{abstract}
\end{frontmatter}

\section{Introduction}
%
%
%
%

\subsection{Context}
\begin{figure*}[h]
\centering
\includegraphics[width=0.9\linewidth]{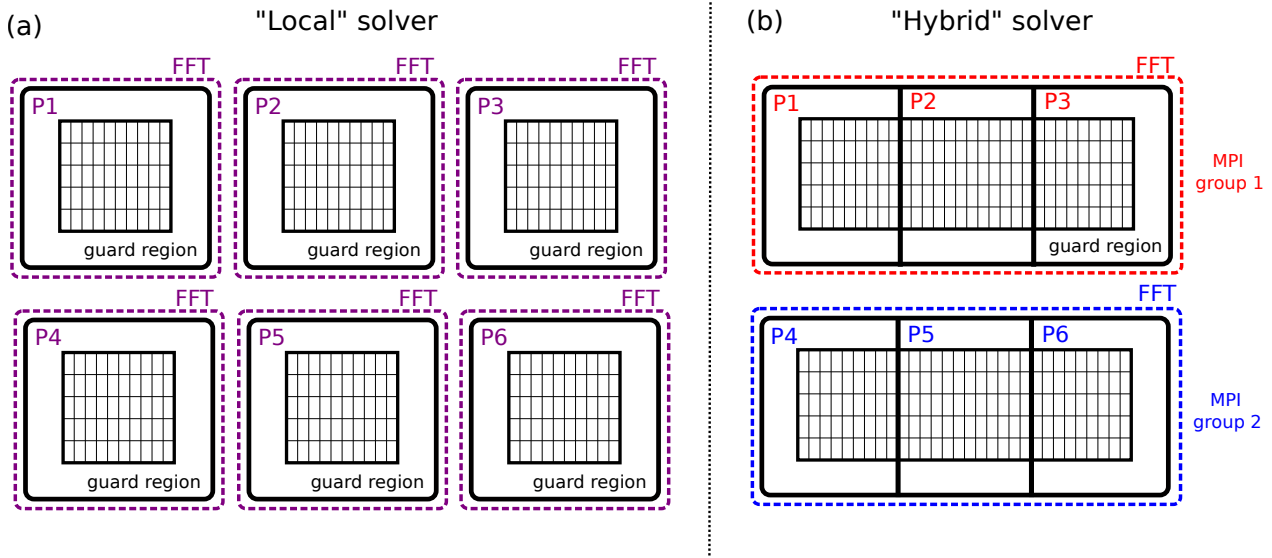}
\caption{Parallelization strategies for pseudo-spectral Maxwell solvers.  (a) is a sketch of the 'Local' approach where the simulation domain is split into multiple sub-domains with guard cells appended at each sub-domain boundary. Guard cells hold copies of electromagnetic fields from adjacent sub-domains. Each sub-domain is handled independently by an MPI process. At each time step: (i) Maxwell's equations are advanced independently on each MPI sub-domain using shared-memory 'local' FFTs and (ii) guard cells are exchanged between adjacent MPI sub-domains. Panel (b) shows a sketch of the new 'hybrid' approach presented in this paper. It consists in grouping several MPI sub-domains into a larger MPI group and perform a distributed FFT on the MPI ranks of this group. Guard cells are solely appended at the boundary of the MPI group leading to less memory redundancy and thus significant memory savings. At each time step: (i) Maxwell's equations are advanced independently on a MPI group using a distributed FFT. (ii) Guard cells are exchanged between MPI groups.}
\label{fig2}
\end{figure*}

The ElectroMagnetic (EM) Particle-In-Cell (PIC) method \cite{HockneyEastwoodBook,Birdsalllangdon} has been the method of choice to model kinetic effects at play in the physics of high intensity laser plasma interactions also known as 'Ultra-High Intensity' (UHI) physics. To describe the plasma and electromagnetic field dynamics, the EM-PIC algorithm self-consistently advances Maxwell's equations on a grid (Maxwell solver) and equations of motion of plasma pseudo-particles. As there is no diagnostic of the plasma and fields evolution at the extremely small time and length scales involved in UHI physics, the EM-PIC method has been crucial to interpret experiments, develop theoretical models as well as propose and guide novel experiments. 

In the last decades, the Maxwell solver used in most EM-PIC codes has been the so-called Finite Difference Time Domain (FDTD) Yee solver \cite{Yee} that operates a second order finite difference in time and space to discretize Maxwell's equations. This method has been very popular since the advent of distributed-memory parallel computers because it can be efficiently parallelized using a standard Cartesian domain decomposition. This parallelization method splits the simulation domain into sub-domains, with guard cells appended at the edges of each sub-domain that stores electromagnetic fields values from immediate neighboring sub-domains. At each time step, Maxwell's equations are then advanced on each processor sub-domain independently and guard cells exchanged between sub-domains. 

FDTD solvers are very local and demonstrated scaling on up to a million cores \cite{fonseca2013exploiting, vincenti201717} as required by the most demanding 3D EM-PIC simulations. Nevertheless, they induce spurious numerical dispersion of electromagnetic waves that reveals highly detrimental in the accurate modeling of laser-plasma-based applications. Mitigation of numerical dispersion errors usually requires very high spatio-temporal resolution that has prevented doing realistic 3D modeling on a large class of problems (including laser-plasma mirror interactions \cite{Blaclard2016,vincenti2018ultrahigh}) for a long time.

In contrast, ultra-high order $p$ (stencil of width $p/2$), and in the infinite order $p\rightarrow\infty$ limit, FFT-based pseudo-spectral solvers, which advance electromagnetic fields in Fourier space (rather than configuration space), can bring much more accuracy than FDTD solvers for a given resolution. In particular, Haber \textit{et al} showed \cite{Habericnsp73} that under weak assumptions, Fourier transforming Maxwell's equations in space yields an analytical solution for electromagnetic fields in time, called the Pseudo-Spectral Analytical Time Domain (PSATD) solver, which is accurate to machine precision for the electromagnetic modes resolved by the mesh. As a consequence this solver enables infinite order, imposes no Courant time step limit in vacuum and has no numerical dispersion. By lowering the resolution needed to reach a required accuracy compared to FDTD solvers \cite{Blaclard2016,vincenti2018ultrahigh,jalas2017accurate}, PSATD-type solvers have the potential to strongly reduce the time-to-solution of a large class of problems. 

\subsection{Scalability limits of global FFT-solvers}

 Nevertheless, pseudo-spectral solvers employing distributed FFTs (later called 'global' FTT solvers in the remainder of this article) have not been popular so far due to the difficulty to scale the distributed FFTs beyond $10,000$ cores \cite{Habibarxiv2012}, which is not enough to take advantage of the largest supercomputers (with up to millions of cores) required for 3D modeling. 
 
 The main barrier to scale FFT computations emerges from the overhead induced by the global communications ('All' to 'All'-type communication) required to transpose the data among all computing units. As a consequence, developing massively parallel distributed FFTs algorithms is extremely challenging and is still an active research area for computer scientists.
 %
 
\subsection{Recent advances in the scaling of FFT-based Maxwell solvers}

To break this scalability barrier, a pioneering grid decomposition technique was recently proposed for pseudo-spectral FFT-based electromagnetic solvers \cite{VayJCP2013}. This technique consists in using a standard Cartesian domain decomposition strategy to parallelize pseudo-spectral solvers (cf. Fig. \ref{fig2}- (a)). Maxwell's equations are solved independently on each MPI processor sub-domain using local FFTs (instead of a single global distributed FFT). Guard cells are exchanged between adjacent sub-domains at each time step. This technique is much more local and can be efficiently scaled providing that the number of guard cells is not too high. It comes however with small stencil truncation errors at sub-domain boundaries when the stencil width $p/2$ is higher than the number of guard cells $n_g$. An important theoretical study \cite{Vincenti2016a} recently derived the analytical expression for the amplitude and phase of these truncation errors. It showed that choosing a very high but finite order $p$ stencil can already strongly reduce truncation errors compared to infinite order while still ensuring extreme accuracy. The model also pointed out that very high order solvers $p>100$ can be used with a moderate number of guard cells ($n_g\ll p/2$) while still guaranteeing low levels of truncation errors (potentially below machine precision). By providing  the number of guard cells required to obtain a given level of truncation error as a function of solver order, time step and mesh size, this model is crucial to enable the parallelization technique in production simulations. 

This new type of FFT-based solvers (later called 'local' FFT-based solvers) has been implemented and optimized in the high performance library PICSAR \cite{vincenti201717,vincenti2018ultrahigh}. They led to very good scaling on up to a million cores even for a moderate ($n_g<10$) number of guard cells and high solver orders $p=100$. They notably enabled the very first accurate 3D simulations of laser-plasma mirror interactions \cite{vincenti2018ultrahigh,vincentiPRL2018,chopineau2018identification} that were used to interpret the latest experimental results at CEA Saclay obtained with the 100TW UHI100 laser.

\begin{figure*}[ht]
\centering
\includegraphics[width=15cm]{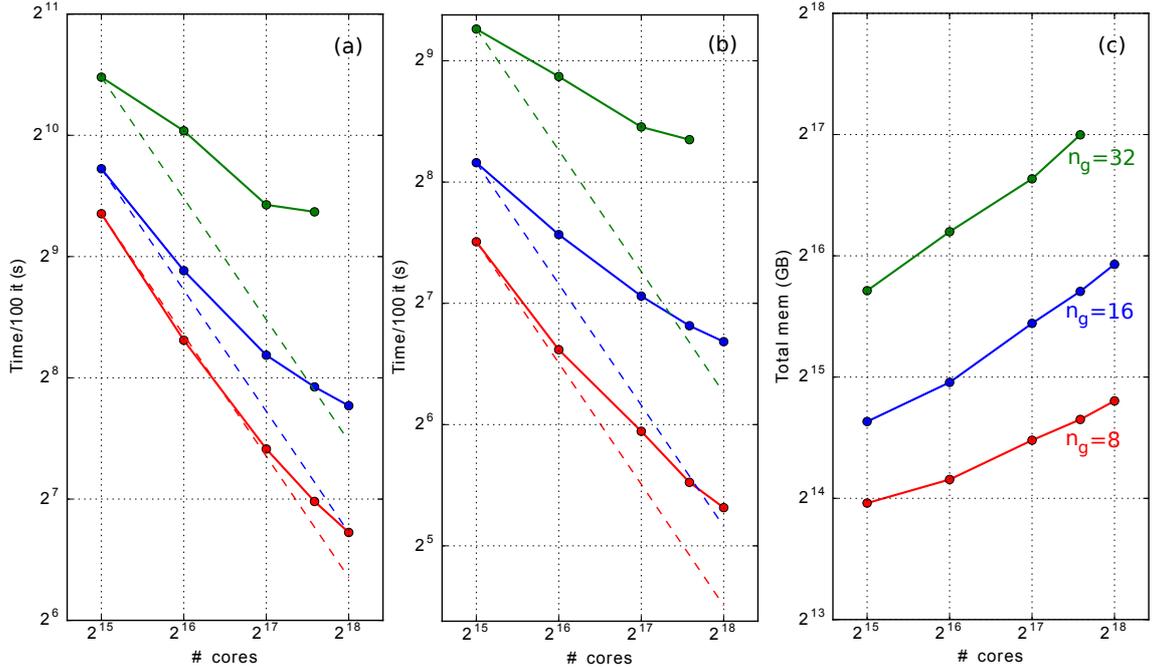}
\caption{Evolution of strong scaling and memory consumption (grid quantities) of the pseudo-spectral PIC algorithm with the number of guard cells $n_g$ used (8 guard cells: red, 16 guard cells: blue, 32 guard cells: green) on  Cray XC40 THETA cluster at ALCF (using  32768-262144 KNL cores) with one OpenMP thread per MPI. The FFTs are computed using the Intel MKL library. The simulation box consists in a homogeneous plasma with one particle per cell for both electrons and ions. Grid size: 240 x 6144 x 12288 grid cells. Panel (a) represents the scaling  of the full PIC loop. Panel (b) represents the scaling of the pseudo-spectral solver only, including MPI-exchanges for grid quantities. Panel (c) represents the total memory consumption of grid quantities (ie field quantities). }
\label{fig1}
\end{figure*}

\subsection{Need for a more general massively parallel FFT-based Maxwell solver}

Yet, local FFT-based solvers may not be adapted to the modeling of certain classes of problems where the level of truncation errors need to be lower than machine precision for instance. In that case, the number of guard cells required $n_g$ might be higher and can strongly affect the strong or weak scaling of the local solver. 

This is illustrated on Fig. \ref{fig1} where we compared the strong scaling of the local solver for different numbers of guard cells. As we can see on panels (a) and (b), although cases with a moderate number of guard cells succeed in keeping a good strong scaling, cases with a  high number of guard cells quickly loose efficiency as the number of processors is increased. We can also note on panel (c) that due to an increased data redundancy, the total volume of memory required grows considerably as the number of guard cells/processors is increased. 

Both limitations in terms of memory consumption and strong scaling of the local solver call for new parallelization strategies allowing for the use of an arbitrarily high number of guard cells. 
In this article, we present a novel massively parallel pseudo-spectral solver that ensures excellent strong/weak scaling at large scale (up to 800k cores) while allowing for the use of high number of guard cells to significantly reduce truncation errors. The remainder of the paper is divided into three additional sections: 
\begin{enumerate}
\item In section 2, we present the principle of the new parallelization method and its implementation in the high performance PIC library PICSAR, 
\item In section 3, we present the benchmarks of the novel method on two large clusters (MIRA and THETA) available at the Argonne Leadership Computing Facility (ALCF), 
\item In section 4, we conclude by presenting the perspectives brought by this novel method for the field of laser-plasma interaction. 
\end{enumerate}

\section{Generalized massively parallel FFT-based Maxwell solver}
In this section, we propose a new parallelization technique for the FFT-based Maxwell solver. This new type of FFT-based solver (later called 'hybrid' FFT-based solver) is a more general approach than the local solver as it permits the use of an arbitrary high number of guard cells while still ensuring extremely good scalability. 

\subsection{Principle of the new solver}

The principle of the hybrid solver is illustrated on Fig. \ref{fig2}- (b). Adjacent MPI sub-domains are grouped into MPI groups (two groups on Fig. \ref{fig2}- (b)). Guard cells are now solely appended to the MPI group boundaries (and not to each MPI sub-domain boundaries). At each time step: (i) Maxwell's equations are advanced independently on a MPI group using a distributed FFT and (ii) guard cells are exchanged between adjacent MPI groups.\\

As we now detail, this solver allows for significant memory savings as well as a reduction of the total volume of data exchanged.\\

\subsection{Advantages in terms of memory}

Let us assume a cubic mesh of size $n_x\times n_y\times n_z=n^3$ splitted into $n_p$ MPI sub-domains along each direction $x$, $y$ and $z$. If $n_g$ guard cells are used for each MPI sub-domain, the total memory occupied by electromagnetic field arrays varies as: 

\begin{equation}
M_{tot}^{loc}=\mathcal{O}\left(n_p^3\times\left[\frac{n}{n_p}+2n_g\right]^3\right)
\end{equation}

As expected, one can notice that this memory strongly increases with the number of guard cells $n_g$. In addition, the maximum number of processors that can be used along each axis for this problem is given by $\frac{n}{n_p}=n_g$, for which the total memory used culminates to: 

\begin{equation}
M_{tot}^{loc}=27 M_{n}
\end{equation}

where $M_{n}$ would be the total memory occupied by field arrays without any extra memory coming from guard cells. This maximum limit $M_{tot}^{loc}$ does not depend on the number of guard cells $n_g$. However, this maximum limit is attained for a much lower number of processors when $n_g \gg 1$, potentially limiting the maximum number of processors that can be used due to memory limitations.\\

Let us now assume that MPI domains are grouped and that we use $n_{mpi}$ MPI processes per group along each axis. In this case, the total memory occupied by electromagnetic field arrays varies as: 

\begin{equation}
M_{tot}^{hyb}=\mathcal{O}\left(\frac{n_p^3}{n_{mpi}^3}\times\left[\frac{n}{n_p}n_{mpi}+2n_g\right]^3\right)
\end{equation}

For $\frac{n}{n_p}=n_g$, we now obtain: 

\begin{equation}
M_{tot}^{hyb}=\left(\frac{2+n_{mpi}}{n_{mpi}}\right)^3 M_{n}
\end{equation}

The memory gain of the hybrid solver compared to local solver for the maximum limit of $\frac{n}{n_p}=n_g$ is thus: 

\begin{equation}
G^{3}=\frac{M_{tot}^{loc}}{M_{tot}^{hyb}}=3^3\left(\frac{n_{mpi}}{2+n_{mpi}}\right)^3
\end{equation}

If we only make MPI groups along $d$ axes ($d\leqslant 3$), this formula becomes: 

\begin{equation}
G^{d}=3^d\left(\frac{n_{mpi}}{2+n_{mpi}}\right)^d
\end{equation}

\begin{figure}[h]
\centering
\includegraphics[width=0.8\linewidth]{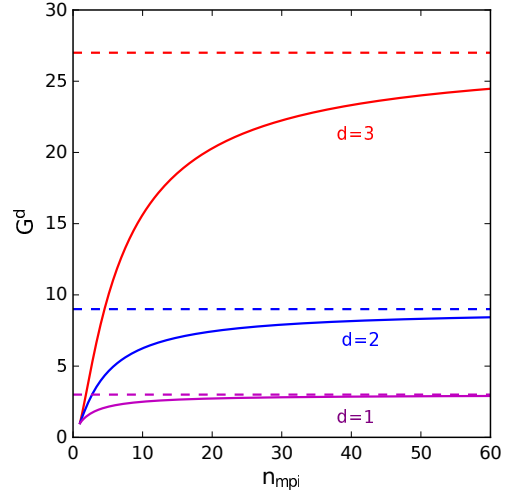}
\caption{Gain $G^d$ in terms of memory of the hybrid approach compared to the local approach (corresponding to $n_{mpi}=1$). The curves represents $G^d$ as a function of the number of MPI sub-domains per group $n_{mpi}$ for d=1 (purple curve), d=2 (blue curve) and d=3 (red curve). Dashed lines represent the asymptotic limit of $G^d$ when $n_{mpi}>>1$.}
\label{fig3}
\end{figure}

The number of axis $d$ along which MPI sub-domains can be grouped directly depends on the number of axis along which the distributed FFT can be parallelized. For the FFTW library \cite{FFTWref} (1D slab decomposition) we can parallelize the distributed FFT along $z$ only (in Fortran) and thus group MPI sub-domains solely along $z$ (i.e $d=1$). For the P3DFFT library \cite{pekurovsky2012p3dfft} (2D pencil decomposition), the distributed FFT can be parallelized along $y$ and $z$ and MPI subdomains can therefore be grouped along two directions (i.e. $d=2$). The gain $G^{d}$ as a function of $n_{mpi}$ is plotted on Fig. \ref{fig3} for different values of $d=1,2,3$. One can notice that for a few $n_{mpi}>10$ per groups, the total gain can already approach its maximum asymptotic value ($G=27$ for $d=3$, $G=9$ for $d=2$ and $G=3$ for $d=1$). 


\subsection{Gains in terms of volumes of guard cells data exchanged}

Similarly, one can estimate the gain in terms of volume of guard cells data exchanged between the hybrid and local approaches in the limit $\frac{n}{n_p}=n_g$:

\begin{equation}
G_{guard}^{d}=\frac{M_{tot}^{loc}-M_n}{M_{tot}^{hyb}-M_n}=\frac{3^d-1}{\left(1+2/n_{mpi}\right)^d-1}
\end{equation}

For instance, the maximum gain on the total volume of guard cell exchanged for $d=2$ (with P3DFFT) and $n_{mpi}=5$ can be as large as $8$. 

\subsection{Advantages in terms of execution time of the FFT}

Another important time gain expected from our hybrid parallelization technique emerges from the execution time of the FFT. In the following, we first estimate the time complexity of the distributed FFT algorithm. In the light of this estimate, we then present the advantages of the hybrid solver compared to the local and global solvers. 

\subsubsection{Estimate of distributed FFTs computation time}

Assessing the complexity of distributed memory FFT is important to understand and take advantage of the scalability of the hybrid solver. Distributed-memory FFTs performance strongly depends on both communication network and computer architecture.
For multi-dimensional FFTs, most distributed-memory FFT libraries follow the same computation scheme detailed below. For each axis A of a 3D array:  
\begin{enumerate}
\item If processors have all data in their local memory along $A$, directly compute FFT along $A$.  
\item If data along $A$ is distributed on different processors, first transpose the 3D array so that each processor have all data along $A$ and then compute FFT along $A$. 
\end{enumerate}

Following this computation scheme, we define the total time to perform a 3D FFT as the sum of the time required to transpose data $T_{tr}$ and the time to compute the FFTs $T_{c}$: 
\begin{equation}
T_{FFT} = T_{c} + T_{tr}
\label{eq:tfft}
\end{equation}
The time complexity of a 3D FFT is known to be:
\begin{equation}
T_{c}  = \alpha  n^3 \log{}n^3
\end{equation}
where n is the global array size along each axis and $\alpha $ a machine-dependent parameter.

For a distributed-memory FFT with pencil decomposition, the data is distributed over $n_{proc} = n_p^2$ processors. In this case assuming perfect scaling, the computation time $T_c$ per processor is: 
\begin{equation}
T_{c}  =  \alpha  \left( \frac{ n^3 \log{n^3} }{n_{proc}}\right)    
\end{equation}

On the other hand, the transpose time $T_{tr}$ is very network depend. Following the same reasoning  as in \cite{pekurovsky2012p3dfft} to estimate the communication time of the transpose operation, we get:
\begin{equation}
T_{tr}  = \beta \left( \frac{ n^3 }{\sigma_{bi}[n_{proc}]}\right) + \gamma \left( \frac{ n^3 }{n_{proc}.\sigma_{mem}}\right)
\label{eq:t_tr_gen}
\end{equation}

where $ \sigma_{bi}(n_{proc})$ is the bisection bandwidth of the network, $\beta$ and $ \gamma$ two machine and network dependent parameters and $\sigma_{mem}$ is the memory bandwidth per MPI task. 

For a large number of MPI tasks, we can neglect the term $ \gamma  \frac{ n^3 }{n_{proc}.\sigma_{mem}} $ compared  to $\beta \frac{ n^3 }{\sigma_{bi}[n_{proc}]}$ in eq (\ref{eq:t_tr_gen}) as the inter-node data transposition is more costly than the intra-node data-transposition (which only requires memory copies).

The bisection bandwidth is a function of the number of processors and depends on the nature of the supercomputer interconnection network.
For a 5D torus network such as the one equipping the IBM BG-Q MIRA cluster at the ALCF, $\sigma_{bi}[n_{proc}]$ should scale as $ {n_{proc}}^{4/5}$. For the case of the THETA cluster (equipped with a Dragon fly network), $\sigma_{bi}[n_{proc}]$ should scale as $ {n_{proc}}$. In practice, we estimated the bisection bandwidth by fitting the global transposition time as a power of $n_{proc}$. For both MIRA and THETA, the best fit found for $\sigma_{bi}[n_{proc}]$ scales as $ {n_{proc}}^{4/5}$. For THETA, this is lower than the expected theoretical value of $n_{proc}$. As opposed to MIRA where compute nodes are allocated contiguously, compute nodes on THETA can be allocated at remote locations on the network, depending on the cluster occupancy at a given time. This might explain the lower-than-expected bisection bandwidth on THETA. Based on the bisection bandwidth estimates, we therefore used the following expression of $T_{FFT}$ for both machines: 

\begin{equation}
T_{FFT}  = \alpha \frac{ n^3 \log{}n}{n_{proc}} + \beta \frac{n^3}{{n_{proc}}^{4/5}}
\label{eq:tfftexpl}
\end{equation}
\begin{figure}[ht]
\centering
\includegraphics[width=\linewidth]{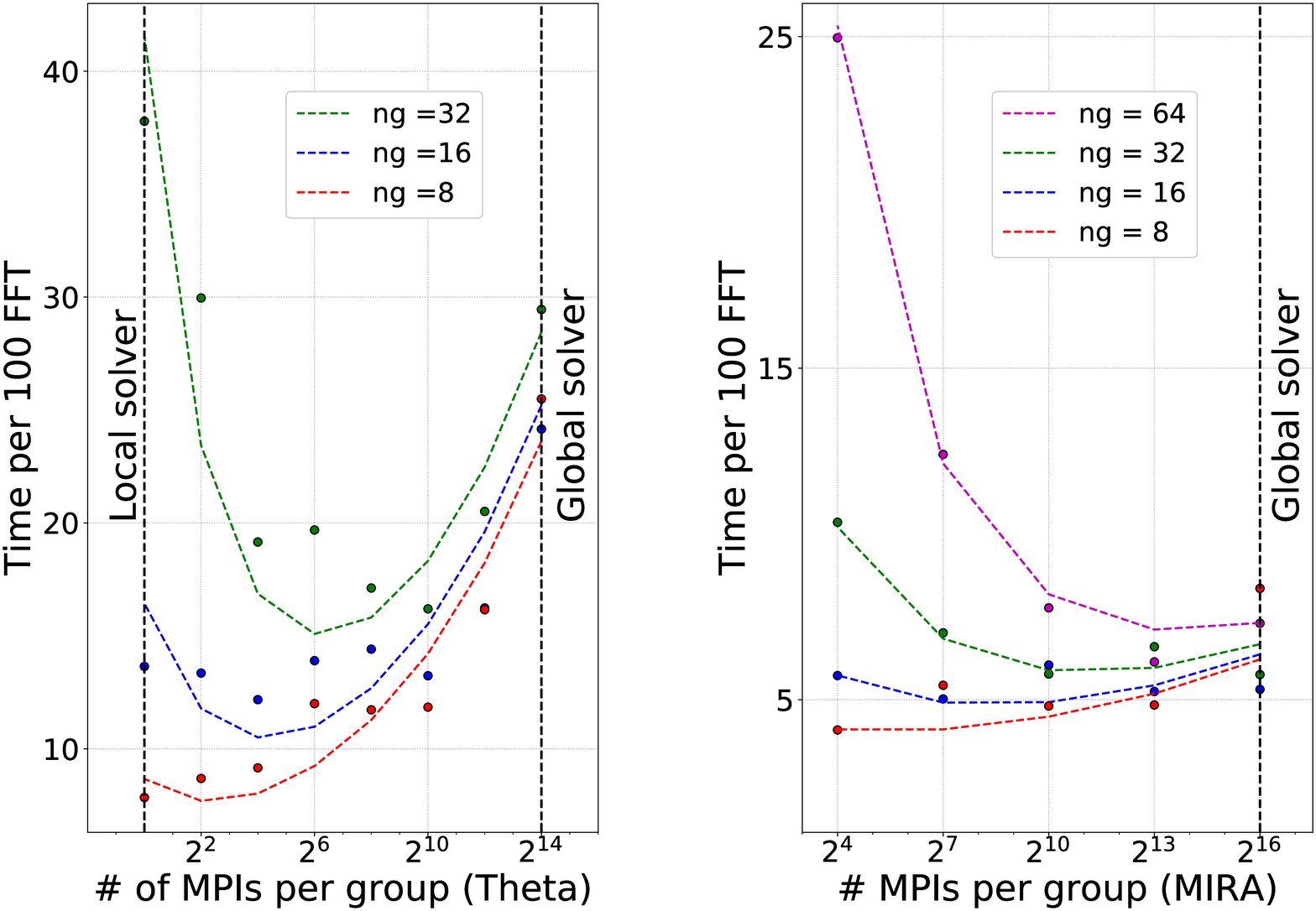}
\caption{FFT total execution time as a function of the total number of MPI processes per group and the number of guard cells. The left panel represents the data from Theta and the right pane represents the data from Mira. Dots represent measured FFT execution time while dashed lines represent the fitting curve following eq \ref{eq:tfftexpl} }
\label{fig8}
\end{figure}

From formula (\ref{eq:tfftexpl}), one can understand why distributed-memory FFTs do not scale well at large scale on massively parallel computer architectures: for a large number of processors and a relatively small array size, the transpose operation will dominate the FFT computation $\alpha \frac{\log{}n }{n_{proc}} \ll  \frac{\beta}{n_{proc}^{4/5}}$ hence resulting in a poor scaling proportional to $1/n_{proc}^{4/5}$. On the other hand, for large scale data set and relatively small number of processors, the computation term of the FFT will dominate the total time and $\alpha \frac{\log{}n }{n_{proc}} \gg  \frac{\beta}{n_{proc}^{4/5}}$ thus resulting in a good scaling proportional to $1/n_{proc}$. 

\subsubsection{Advantages of the hybrid solver over local and global solvers}

In the light of eq (\ref{eq:tfftexpl}), one can note that our hybrid solver makes it possible to reduce the relative weight of the transpose operation time $T_{tr}$ in the total FFT time $T_{FFT}$  by properly tuning the number of MPI processes per group $n_{mpi}$ on which the distributed FFT is performed. This should lead to a better performance than purely global solvers in most cases.

In addition, the hybrid solver should also outperform the local solver at large scale. Indeed, one can show that the execution time for the FFT (neglecting guard cell exchanges) for the local solver scales as: 

\begin{equation}
T_{FFT}^{loc}  \propto n\left[\frac{n}{n_p}+2n_g\right]^2 \log{} n\left[\frac{n}{n_p}+2n_g\right]^2
\label{eq:tfftloc}
\end{equation}
where we assumed a 2D domain decomposition. For a large number of processors $n_p$ such as $n/n_p \rightarrow n_g$, the execution time $T_{FFT}^{loc}$ becomes constant and the parallel efficiency of the local solver drops considerably. In contrast, for the Hybrid solver this execution time should write: 
\begin{equation}
T_{FFT}^{hyb}  \propto \frac{n}{n_{mpi}^2}\left[\frac{n}{n_p}n_{mpi}+2n_g\right]^2 \log{} \frac{n}{n_{mpi}^2}\left[\frac{n}{n_p}n_{mpi}+2n_g\right]^2
\label{eq:tfftloc}
\end{equation}
where we assumed a 2D pencil decomposition and a negligible transpose time compared to the FFT computation time.  Choosing large enough MPI groups such that $n/n_{p}n_{mpi}\gg n_g$ therefore leads to: 
\begin{equation}
T_{FFT}^{hyb}  \propto \frac{n^2}{n_{p}^2} \log{} \frac{n^2}{n_{p}^2} 
\label{eq:tfftloc}
\end{equation}
which would ensure very good scaling of the hybrid solver at large scale. 

\subsubsection{Benchmarks of the hybrid solver vs local and global solvers}
To demonstrate the superiority of the hybrid solver over local and global solvers, many 3D simulations were run on both THETA and MIRA where the number of guard cells $n_{g}$ and MPI processes per groups $n_{mpi}$ were varied (cf. Fig. \ref{fig8}). In all these simulations, we used the P3DFFT library (pencil decomposition) to perform the distributed FFTs allowing to group $n_{mpi,y}$ and $n_{mpi,z}$ MPI processes along the $y$ and $z$ directions. Below are given details of the 3D cases run on MIRA and THETA: 

\begin{itemize}
\item \textbf{On MIRA}:  the array size per direction was chosen to $n=2048$ and the number of guard cells $n_g=8$, $16$ and $64$. $8192$ BG/Q nodes were used with $8$ MPI processes per node (total of $2^{16}$ MPI tasks). For each case, the total number of MPI processes per group $n_{mpi,y}\times n_{mpi,z}$ was varied as follows: $2^4$, $2^7$, $2^{10}$, $2^{13}$ and $2^{16}$, with $n_{mpi,y}\times n_{mpi,z}=2^{16}$ corresponding to the global solver and $n_{mpi,y}\times n_{mpi,z}=1$ to the local solver.

\item \textbf{On THETA}: the array size was chosen to $ 512 \times 4096 \times 4096 $ cells and the number of guard cells $n_g = 8$, $n_g =16$, $ n_g =32$. 512 KNL nodes were used with 64 MPI per node for a total of $32768$ MPI tasks.  Two MPI tasks were used along the $x$ direction. The rest of the $16384$ MPI tasks were split equally along y and z. The number of MPI processes per group along y and z direction was varied from $n_{mpi,y}\times n_{mpi,z}=2^{14}$ (global solver), to $n_{mpi,y}\times n_{mpi,z}=1$ (local solver). 

\end{itemize}

For each simulation, we collected the averaged execution time of FFTs among all MPI ranks. We chose to use the averaged time between these ranks since the problem was very well load-balanced between MPI processes (standard deviation of execution time did not exceed $10$\% among all MPI ranks). These execution times are displayed using colored markers on Fig. \ref{fig8}. From these execution times, we could estimate the values of the parameters $\alpha$ and $\beta$ in the theoretical expression of $T_{FFT}$ (cf. eq (\ref{eq:tfftexpl})) using a least square algorithm. On MIRA, we obtained $ \alpha = 1.$ and $ \beta = 1.5414$. On THETA, we obtained $ \alpha = 1.$ and $ \beta = 67.1$  We checked that using different box sizes and other simulation parameters led to similar results on both Theta and  MIRA. One can see on Fig. \ref{fig8} (cf. dashed lines) that the least square fit obtained leads to an acceptable matching between expected and measured timings.

Fig. \ref{fig8} shows the optimal number of MPI processes per group that minimizes the total execution time increases with the number of guard cells. For a low number of guard cells $n_g=8$, a few MPI processes per group (around 8) seems the best approach whereas for a higher number of guard cells $n_g=16$, $20$ MPI processes per group are required. As a consequence, having a good guess about the values of $\alpha $ and $\beta$ on a given network/computer architecture can help tune the hybrid solver to take advantage of the performance boost of our hybrid solver. The procedure to find the optimal number of groups for given domain decomposition/cluster architecture will be automated and added to the PICSAR library. 

Although the model of eq. (\ref{eq:tfftexpl}) is quite simple and omits various operations involved in the FFT computation (data transfer among one compute node / data transposition among one MPI task ...), the error between the fit and the measured timing does not exceed $22$\%. This can also be explained by the the fluctuation coming from the 1D FFT computation since the $ n {} \log {} n$ scaling is not perfect. 

Further tuning can be done on different machines to predict the optimal number of MPI processes per group. This optimum lies between $1$ (global solver) and $n_{proc}$ (local solver), depending on the machine and the problem size. We have noted that MIRA can support rather large MPI groups while keeping a good scaling, whereas on THETA, our method gives better results with smaller groups. For critical cases where $ \frac {n}{n_p} = n_g $ where $n_g$ is the number of guard cells, we have noticed that it is always better to use the hybrid solver since it gives a substantial gain in terms of performance and memory saving.

\subsection{Coupling of the hybrid Maxwell solver with the PIC algorithm}

\subsubsection{Brief reminder of the PIC algorithm}

\begin{figure}[h]
\centering
\includegraphics[width=0.9\linewidth]{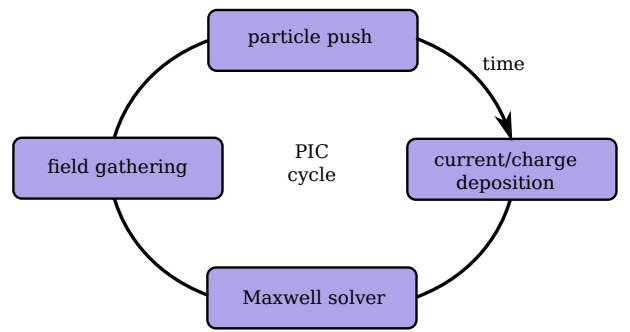}
\caption{Sketch of the PIC cycle.}
\label{fig4}
\end{figure}

The PIC algorithm self-consistently resolves Maxwell's equations on a grid and motion equations of plasma pseudo-particles . The successive steps of the PIC cycle are sketched on Fig. \ref{fig4}. At each time step: 
\begin{itemize}
\item particles are advanced using equations of motion (particle push) knowing the values of the electromagnetic fields at their position, 
\item once particles are advanced, current/charge contribution of each particle on the grid is deposited using linear or higher order interpolation, 
\item once electromagnetic sources are known on the grid, Maxwell's equations are solved to advance electromagnetic fields in time and space (Maxwell solver), 
\item once electromagnetic fields are known at next time step, field values are gathered from the mesh to particle's position using interpolation, usually at the same order as the deposition. 
\end{itemize}

\subsubsection{Dual grid decomposition for efficient load balancing}

\begin{figure}[h]
\centering
\includegraphics[width=1\linewidth]{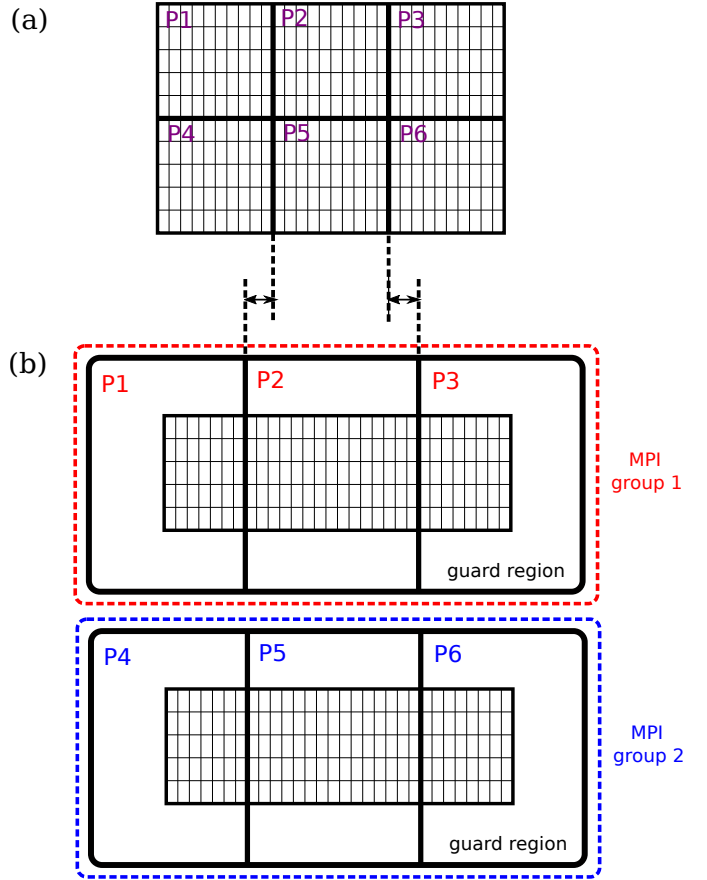}
\caption{Dual domain decompositions used to load balance (a) particle-mesh/particle operations of the PIC loop (b) FFT computations of the Maxwell solver step. }
\label{fig5}
\end{figure}

All the difficulty in coupling the hybrid solver with the PIC algorithm lies in the efficient load balancing of (i) the particle and particle-mesh operations (particle push, field gathering, current/charge deposition) on the one hand and (ii) the Maxwell solver on the other hand. This has been illustrated on panels (a) and (b) of Fig. \ref{fig5}.\\

Panel (a) shows the domain decomposition $D_1$ used to efficiently load balance particle and particle/mesh operations (plasma is assumed to be homogeneous). Limits of MPI sub-domains are highlighted using black solid lines. Guard cells required for particle-mesh operations (deposition/gathering) and of width equal to the order of deposition/gathering have not been represented for more clarity. 

Panel (b) shows the domain decomposition $D_2$ that would be required to efficiently load balance the FFTs in the Maxwell solver step. One can see that in that case, the limits of MPI sub-domains in $D_1$ do not coincide with the limits of MPI sub-domains in $D_2$ due to the presence of guard cells appended at the MPI group boundaries and needed in the computation of FFTs.

Load balancing all steps of the PIC loop thus requires keeping two different grids: (i) one grid $G_1$ for the particle and particle-mesh operations (ii) one grid $G_2$ for the FFTs. Note that the additional grid $G_1$ is also usually employed for smoothing \cite{GODFREY20141} or mesh refinement \cite{vay2018warp} in the PIC algorithm. This comes with a slight additional memory cost ($\approx 30\%$) that yet still allows for large total memory gain thanks to the decrease of guard cells data redundancy with the hybrid solver.  

We now detail how the coupling between the two grids is managed in the PIC cycle. At each time step:
\begin{enumerate}
\item Fields from $G_1$ are copied to $G_2$ (including copies of data to the guard cells of MPI groups in $G_2$). Overlapping grid regions pertaining to same MPI domains on $G_1$ and $G_2$ are simply copied from $G_1$ to $G_2$. Other regions of $G_2$ are updated using MPI exchanges of grid data from $G_1$ to $G_2$, 
\item Maxwell's equations are solved using the hybrid solver on $G_2$, 
\item After the Maxwell solve step, field data (without guard cells) are copied from $G_2$ to $G_1$. Overlapping grid regions pertaining to same MPI domains on $G_2$ and $G_1$ are simply copied from $G_2$ to $G_1$. Other regions of $G_1$ are updated using MPI exchanges of grid data from $G_2$ to $G_1$. 
\end{enumerate}
Note that this implementation does not actually require direct exchanges of guard cell data between MPI groups as mentioned in the previous section. Guard cells of $G_2$ are instead directly filled from $G_1$ during the first step of the coupling.  

\begin{figure*}[ht]
\centering
\includegraphics[width=0.7\linewidth]{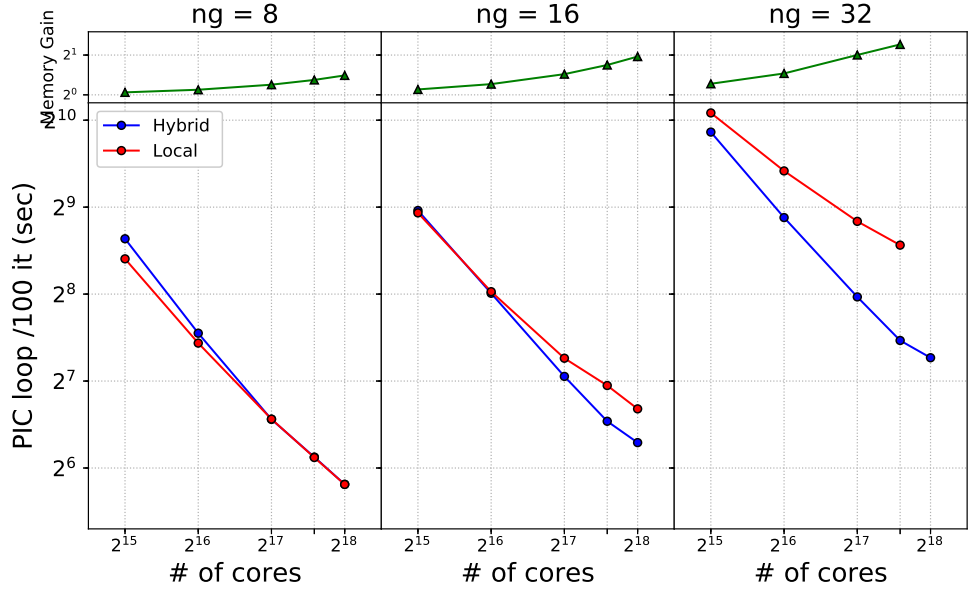}
\caption{Local vs Hybrid + slab decomposition on THETA. Each panel represents a different number of guard cells.The blue curves stand for the hybrid solver data, while the red curves stand for the local solver data. The green line represents the total memory gain brought by the hybrid solver compared to the local solver. 
}
\label{fig_pic_fftw_theta}
\end{figure*}

\subsection{Implementation in the PICSAR library}
 The \textbf{P}article-\textbf{I}n-\textbf{C}ell \textbf{S}calable \textbf{A}pplication \textbf{R}esource (PICSAR) \cite{Vincenti2017, vincenti201717} is an open-source high performance library intended to help scientists porting their code to the next generation of exascale computers. PICSAR contains highly optimized PIC routines exploiting the three levels of parallelism that modern architectures offer (Internode parallelism, Intranode parallelism, Vectorization) as well as optimized parallel I/O routines. 

The hybrid Maxwell solver has been fully implemented in PICSAR and can currently be used by the WARP \cite{VayCSD12, Warp} and WARPX \cite{vay2018warp} codes. Note that the WARPX code also implements a slightly different version of the dual domain decomposition presented in this paper, which won't be presented here.  

PICSAR is written in Fortran 90 and can be compiled in three different modes: 
\begin{itemize}
\item As a Python module that can be directly coupled with other codes also using python as the outermost software layer. Right now, PICSAR is coupled to the WARP PIC code through this layer, 
\item As a static/dynamic library to be used by other Fortran/C/C++ codes. For instance, the WARPX and SMILEI \cite{derouillat2018smilei} PIC codes are presently coupled to PICSAR by this means, 
\item As a self-consistent Fortran 90 PIC code. All tests performed in this paper have been done through this mode.  
\end{itemize}

To perform the distributed FFTs, PICSAR can use either the FFTW or P3DFFT libraries: 
\begin{itemize}
\item FFTW is a well-established GPL FFT library that can  perform shared-memory/distributed FFTs . The distributed FFT only supports parallelization along one axis (slab decomposition), which is the last axis or z-axis in Fortran.
\item P3DFFT is an open source library that can perform distributed FFTs using a pencil domain decomposition. By offering parallelization over two axis, P3DFFT therefore offers more flexibility than the distributed version of FFTW.  To compute the FFTs locally, P3DFFT makes use of existing shared-memory versions of other FFT libraries (currently it only supports FFTW, ESSL).
\end{itemize}

Note that on Intel architectures, such as THETA, the MKL library can be used instead of FFTW through the FFTW-MKL wrapper, which allows performing DFT computations with the MKL library without modifying FFTW directions. All performance tests performed on THETA whether using FFTW or P3DFFT have been done with MKL-FFTW support for shared-memory FFTs, as this proved to bring better performance for both local and hybrid solver. On MIRA, P3DFFT (with FFTW support for shared-memory FFTs) and FFTW were used. To use the hybrid solver, the user needs to specify some additional simulation parameters related to the hybrid solver. These parameters are : 
\begin{itemize}
\item the FFT library used for distributed FFTs (FFTW or P3DFFT), 
\item the number of groups along y and z direction (in the z-direction only when using FFTW),
\item the number of group guard cells along y and z direction (in the -direction only  when using FFTW),
\item an additional flag to enable/disable the use of  "optimized communication strategy" to compute FFTs (Only available in 3D). The "optimized communication strategy"  skips the last data transposition after each FFT in order to save computational time. Indeed the last transposition is only required to have the same data layout for the output array and the input array. This saves computational time but results in a different layout of field arrays in the Fourier space that needs to be taken into account when solving Maxwell's equations. This has been included in PICSAR that benefits from this optimization for both P3DFFT and FFTW modes in 3D simulations.
\end{itemize}

\begin{figure*}[ht]
\centering
\includegraphics[width=0.7\linewidth]{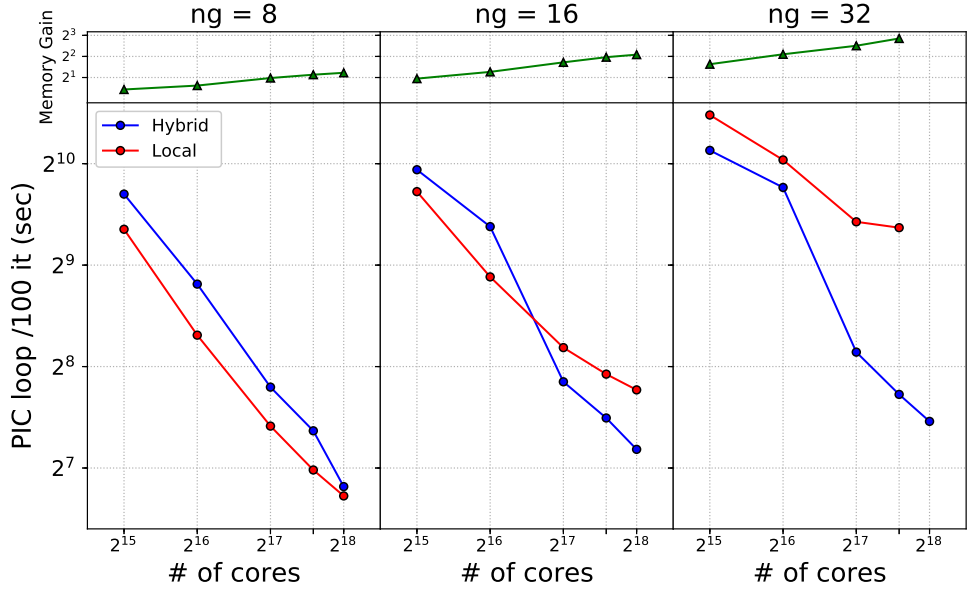}
\caption{Local vs Hybrid + pencil decomposition on THETA. Each panel represents a different number of guard cells. The blue curves stand for the hybrid solver data, while the red curves stand for the local solver data. The green line represents the total memory gain brought by the hybrid solver compared to the local solver. 
}
\label{fig_pic_p3d_theta}
\end{figure*}

Besides, note that when using the hybrid solver, PICSAR calculates the MPI groups topology and creates $n_{group}$ MPI sub-communicators, each sub-communicator being appended to a unique group. The data sizes on each MPI task is set by the FFT library, which results in a new domain decomposition related to the Maxwell's equations solve as illustrated on Fig. \ref{fig5} (b). Finally, FFT plans are initialized by the FFT library on each group. 

The intersection between the two domain decompositions is computed before the PIC loop starts, and a data exchange protocol is determined according to the grid overlaps. If dynamic load-balancing is enabled for the simulation, then this pre-processing step needs to be recomputed accordingly. In practice, the data exchange step is rather computationally cheap and does not add an important overhead to the PIC loop since the majority of data exchanges are simple data copies within the same MPI process. In PICSAR, the data exchange protocol is done in a way that allows both blocking and non-blocking MPI exchanges.

Finally, note that the PICSAR FFT-based Maxwell solver now also supports absorbing boundary conditions implemented using the two-step Perfectly Matched Layers algorithm recently developed in \cite{shapoval2019two}.

\section{Benchmark of the new solver at very large scale}

The new solver has been benchmarked on both THETA and MIRA at very large scale. All benchmarks considered the simulation of a 3D homogeneous plasma with 1 pseudo-particle per cell for both electrons and ions. These benchmarks are presented below. 

\subsection{Benchmark on THETA}
The hybrid solver on THETA has been benchmarked using both P3DFFT and FFTW\_MPI libraries.
On THETA, the FFTW-MKL wrapper has been used with FFTW\_MPI and P3DFFT. 

\begin{figure*}[ht]
\centering
\includegraphics[width=0.7\linewidth]{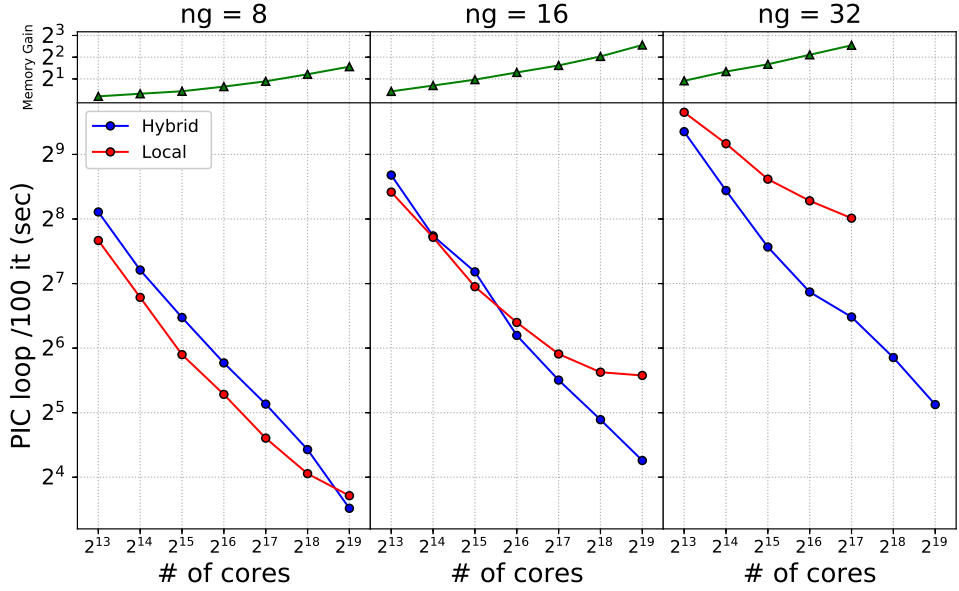}
\caption{Local vs Hybrid + pencil decomposition on MIRA.  Each panel represents a different number of guard cells. The blue curves stand for the hybrid solver data, while the red curves stand for the local solver data. The green line represents the total memory gain brought by the hybrid solver compared to the local solver. 
}
\label{fig_pic_p3d_mira}
\end{figure*}

\subsubsection{Benchmark with the slab decomposition}

The box size is $ n_x \times n_y \times n_z = 160 \times 160 \times 393216$ cells, where larger array sizes were used along z to investigate strong scaling for a large number of processors. Multiple simulations were run with different numbers of guard cells  $ n_g = 8, 16, 32$. The number of KNL nodes used was $N_{nodes}= 512, 1024, 2048, 3072, 4096$ with 64 MPI tasks per node and one OpenMP thread per MPI task. For each simulation we used $n_{mpi}=32$ MPI processes per group.

As shown on Fig. \ref{fig_pic_fftw_theta}, the slab decomposition allows a memory and performance gain of order $\times 2.5$ against the local solver, for a high number of guard cells. The strong scaling efficiency of the hybrid solver for $32$ guard cells is $87\%$ between $512$ and $3072$ KNL nodes, while the local solver efficiency is only about  $47\%$. Besides, note that the hybrid solver allows to run simulations on more nodes than the local solver which is limited by $ \frac{n}{n_p}=n_g$ (note that the last point of the $32$ guard cells simulations with the local solver is missing. 

\subsubsection{Benchmark with the pencil decomposition}

In this benchmark the grid size was $ n_x \times n_y \times n_z= 240 \times 6144 \times 12288$, where a larger size was chosen along $y$ compared to the slab decomposition. We kept the same number of guard cells as for the slab benchmark ($n_g = 8,16,32$). The number of MPI processes per group was fixed to $64$. We used $64$ MPI processes per node and one OpenMP thread/MPI task for all the tests. 

Results of this benchmark are displayed on Fig. \ref{fig_pic_p3d_theta}. Here again, the hybrid solver performs much more efficiently than the local solver at large scale (especially for a large number of guard cells) both in terms of time-to-solution (up to $\times 4$ speed-up) and memory used (up to $\times 8$ less memory). 

\subsection{Benchmark on MIRA}

On MIRA,the hybrid solver was benchmarked using the pencil decomposition only. The grid size was $n_x \times n_y \times n_z = 256 \times 2048 \times 2048 $ cells. All the simulations were ran with $4$ OpenMP threads per MPI process. Each MPI group was composed of $256$ MPI tasks. The number of nodes was varied between $512$ to $16384$.  

This benchmark (cf. Fig \ref{fig_pic_p3d_mira}) shows again a very good strong scaling of the hybrid solver and a very good parallel efficiency at large scale. Moreover, we show that MPI exchanges involved in the PIC loop (apart from the global exchanges involved in the FFT transpose) are less costly when using the hybrid solver (due to a decrease of the total volume of MPI exchanges).

\section{Conclusion}

We presented a novel massively parallel technique for ultra-high order FFT-based Maxwell solvers. As opposed to the previously developed 'local' technique, this 'hybrid' technique (which performs distributed FFTs on groups of neighbouring MPI processes) is very general and allows a very good strong scaling for an arbitrarily high number of guard cells. For large numbers of guard cells, it notably increases the maximum number of MPI processes that can be used to parallelize computations. Besides, by reducing data redundancy, it also has huge benefits in terms of memory savings compared to the 'local' technique for a given problem size. Both the improvements in strong scaling and memory savings will enable larger 3D PIC simulations than previously accessible with the 'local' technique. 

 \section*{Acknowledgements}
The authors would like to thank  R\'emi Lehe and Julien Derouillat for fruitfull discussions. 

This research was supported by the Exascale Computing Project (ECP), Project Number: 17-SC-20-SC, a collaborative effort of two DOE organizations – the Office of Science and the National Nuclear Security Administration – responsible for the planning and preparation of a capable exascale ecosystem – including software, applications, hardware, advanced system engineering, and early testbed platforms – to support the nation’s exascale computing imperative.

This work was supported in part by the Director, Office of Science, Office of High Energy Physics, U.S. Dept. of Energy under Contract No. DE-AC02-05CH11231.

An award of computer time (PICSSAR\_INCITE) was provided by the Innovative and Novel Computational Impact on Theory and Experiment (INCITE) program. This research used resources of the Argonne Leadership Computing Facility, which is a DOE Office of Science User Facility supported under Contract DE-AC02-06CH11357.

\bibliographystyle{unsrt}
\bibliography{references}

\begin{thebibliography}{10}

\bibitem{HockneyEastwoodBook}
R~W Hockney and J~W Eastwood.
\newblock 1988.

\bibitem{Birdsalllangdon}
C~K Birdsall and A~B Langdon.
\newblock Adam-Hilger, 1991.

\bibitem{Yee}
Ks~Yee.
\newblock {\em Ieee Trans. Ant. and Prop.}, 14:302--307, 1966.

\bibitem{fonseca2013exploiting}
Ricardo~A Fonseca, Jorge Vieira, Frederico Fi{\'u}za, Asher Davidson, Frank~S
  Tsung, Warren~B Mori, and Lu{\i}s~O Silva.
\newblock Exploiting multi-scale parallelism for large scale numerical
  modelling of laser wakefield accelerators.
\newblock {\em Plasma Physics and Controlled Fusion}, 55(12):124011, 2013.

\bibitem{vincenti201717}
Henri Vincenti, Mathieu Lobet, Remi Lehe, Jean-Luc Vay, and Jack Deslippe.
\newblock 17 pic codes on the road to exascale architectures.
\newblock {\em Exascale Scientific Applications: Scalability and Performance
  Portability}, page 375, 2017.

\bibitem{Blaclard2016}
G~Blaclard, H~Vincenti, R~Lehe, and JL~Vay.
\newblock Pseudospectral maxwell solvers for an accurate modeling of doppler
  harmonic generation on plasma mirrors with particle-in-cell codes.
\newblock {\em Physical Review E}, 96(3):033305, 2017.

\bibitem{vincenti2018ultrahigh}
Henri Vincenti and Jean-Luc Vay.
\newblock Ultrahigh-order maxwell solver with extreme scalability for
  electromagnetic pic simulations of plasmas.
\newblock {\em Computer Physics Communications}, 228:22--29, 2018.

\bibitem{Habericnsp73}
I~Haber{\it, et al.,}.
\newblock {\em Proc. Sixth Conf. Num. Sim. Plasmas}, pages 46--48, 1973.

\bibitem{jalas2017accurate}
S{\"o}ren Jalas, Irene Dornmair, R{\'e}mi Lehe, Henri Vincenti, J-L Vay, Manuel
  Kirchen, and Andreas~R Maier.
\newblock Accurate modeling of plasma acceleration with arbitrary order
  pseudo-spectral particle-in-cell methods.
\newblock {\em Physics of Plasmas}, 24(3):033115, 2017.

\bibitem{Habibarxiv2012}
S.~Habib{\it, et al.,}.
\newblock {\em arXiv.org}, page 1211.4864.

\bibitem{VayJCP2013}
J.-L. Vay, I.~Haber, and B.~B. Godfrey.
\newblock {\em J. Comput. Phys.}, 243:260--268, 2013.

\bibitem{Vincenti2016a}
H.~Vincenti and J.-L. Vay.
\newblock {\em Comput. Phys. Comm.}, 200:147--167, 2016.

\bibitem{vincentiPRL2018}
H~Vincenti.
\newblock Achieving extreme light intensities using relativistic plasma
  mirrors.
\newblock {\em arXiv preprint arXiv:1812.05357}, 2018.

\bibitem{chopineau2018identification}
L~Chopineau, A~Leblanc, G~Blaclard, A~Denoeud, M~Th{\'e}venet, JL~Vay,
  G~Bonnaud, Ph~Martin, H~Vincenti, and F~Qu{\'e}r{\'e}.
\newblock Identification of coupling mechanisms between ultraintense laser
  light and dense plasmas.
\newblock {\em arXiv preprint arXiv:1809.03903}, 2018.

\bibitem{FFTWref}
\url{http://www.fftw.org/}.

\bibitem{pekurovsky2012p3dfft}
Dmitry Pekurovsky.
\newblock P3dfft: A framework for parallel computations of fourier transforms
  in three dimensions.
\newblock {\em SIAM Journal on Scientific Computing}, 34(4):C192--C209, 2012.

\bibitem{GODFREY20141}
Brendan~B. Godfrey and Jean-Luc Vay.
\newblock Suppressing the numerical cherenkov instability in fdtd pic codes.
\newblock {\em Journal of Computational Physics}, 267:1 -- 6, 2014.

\bibitem{vay2018warp}
J-L Vay, A~Almgren, J~Bell, L~Ge, DP~Grote, M~Hogan, O~Kononenko, R~Lehe,
  A~Myers, C~Ng, et~al.
\newblock Warp-x: A new exascale computing platform for beam--plasma
  simulations.
\newblock {\em Nuclear Instruments and Methods in Physics Research Section A:
  Accelerators, Spectrometers, Detectors and Associated Equipment}, 2018.

\bibitem{Vincenti2017}
H.~Vincenti, M.~Lobet, R.~Lehe, R.~Sasanka, and J.-L. Vay.
\newblock {\em Comput. Phys. Comm.}, 210:145--154, 2017.

\bibitem{VayCSD12}
J.-L. Vay, D~P Grote, R~H Cohen, and A~Friedman.
\newblock {\em Computational Science and Discovery}, 5(1):014019 (20 pp.),
  2012.

\bibitem{Warp}
\url{http://blast.lbl.gov/blast-codes-warp}.

\bibitem{derouillat2018smilei}
Julien Derouillat, Arnaud Beck, F~P{\'e}rez, T~Vinci, M~Chiaramello, A~Grassi,
  M~Fl{\'e}, G~Bouchard, I~Plotnikov, N~Aunai, et~al.
\newblock Smilei: a collaborative, open-source, multi-purpose particle-in-cell
  code for plasma simulation.
\newblock {\em Computer Physics Communications}, 222:351--373, 2018.

\bibitem{shapoval2019two}
Olga Shapoval, Jean-Luc Vay, and Henri Vincenti.
\newblock Two-step perfectly matched layer for arbitrary-order pseudo-spectral
  analytical time-domain methods.
\newblock {\em Computer Physics Communications}, 235:102--110, 2019.

\end{thebibliography}

\end{document}